\documentclass[aps,prb,twocolumn,preprintnumbers,amsmath,amssymb]{revtex4}
\usepackage{graphicx}
\usepackage{xcolor}
\begin{document}

\title{On melting of Boltzmann system of quantum hard spheres.}
\author{S.~M.~Stishov}
\email{stishovsm@lebedev.ru}
\affiliation{P. N. Lebedev Physical Institute, Leninsky pr., 53, 119991 Moscow, Russia}

\begin{abstract}
Melting of a quantum system of hard spheres has been considered in the case when the effects of Bose and Fermi statistics can be neglected. It has been found that the quantum melting line always differs from the classical line with exception for T=0, P=0, where the both lines crossed. It is shown that the classical limit is not reachable at any finite temperatures.
\end{abstract}

\maketitle

\begin{figure}[htb]
\includegraphics[width=40mm]{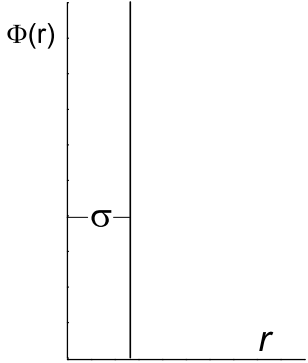}
\caption{\label{fig1} Hard sphere interaction potential .}
\end{figure}

At sufficiently high temperatures, or in  systems with a strong repulsive interaction when particle exchanges are practically impossible, effects of Bose and Fermi statistics can be neglected. However, the system may still be strongly quantum mechanical. The point is that effects of quantum statistics decrease exponentially with temperature increasing, while the "diffraction effects" associated with the wave nature of particles, vanish only as an inverse power of temperature at $T\rightarrow \infty$. So,  there is a significant temperature range in quantum system of hard spheres, where the effects of quantum statistics  play only a minor role~\cite{runge}.
At first, a few words about crystallization of the classical systems of hard spheres. The classical system of hard
of spheres is the simplest non-trivial system with interaction in the form (Fig.~\ref{fig1}),
\begin{equation}
\label{eq:1}
\begin{aligned}
\Phi(r)=0,\ r>\sigma \\    
\Phi(r)= \infty,\ r<\sigma
\end{aligned}
\end{equation}

As was found in the pioneering numerical calculations ~\cite{wood, alder} the hard sphere classical system is able to crystallize and melt by a first order phase transition. It is not difficult to show that the classical hard sphere melting line has the form ~\cite{stishov},
$$P= kT/C,\eqno(2)$$
where P and T are pressure and temperature, C is a constant with volume dimension. As  evident, the only constant with a volume dimension in a hard sphere system is the hard sphere volume. Then for the melting of a classical hard sphere system one may write down an expression (3) with the  dimensionless constant, borrowed from the extensive study ~\cite{hoover}.
     $$P= 11.7kT/\sigma^{3},\eqno(3)$$
 
where  $\sigma$ is a hard sphere diameter. 
 
Hence, as it obvious the melting line (3) is a straight line starting at the coordinate origin $P=O,  T=O$.

Now let us see how quantum effects influence the hard sphere melting. It needs to say that in a dense hard sphere system particles are confined in some sort of a cage, arising due to their impenetrability. Thus, the particles become distinguishable and hence ran by the Boltzmann statistics. So the only quantum feature of the hard sphere system is wave properties of particles. It was shown that quantum effects in the system of hard spheres in the first approximation can be accounted by making use an effective diameter of particles increased by about the de Broglie thermal wave length $\lambda_{T}$  ~\cite{jan, gib, runge} ($\lambda_{T}=\frac{h}{(2\pi mT)^{1/2}}$).
This is expected because quantum particles tend to repel each other according to the uncertainty principle. The effective diameter of quantum hard sphere becomes $\sigma+\lambda/2\sqrt{2}$.
Then one can rewrite (3) in the form,
$$P=11.7 kT/(\sigma +\lambda/2\sqrt{2})^{3}, \eqno(4)$$  
or
$$P=11.7 kT/\sigma^{3}(1+3\lambda/\sigma 2\sqrt{2}), \eqno(5)$$

\begin{figure}[htb]
\includegraphics[width=80mm]{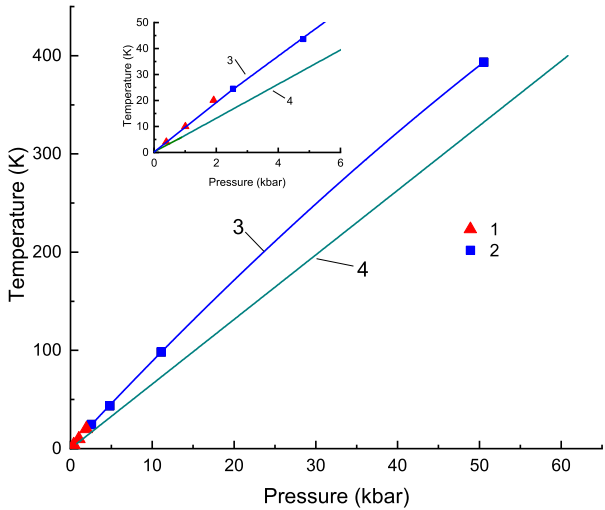}
\caption{\label{fig2}Melting temperatures of quantum and classical hard sphere systems. 1-~Ref.\cite{runge}, 2-~Ref.\cite{sese}, 3-power fit of data~Ref.\cite{runge} , 4-classical melting line. See expressions (6)and (7) in the text. Absolute values of temperature and pressure were obtained from reduced data of Ref.\cite{runge,sese} .}
\end{figure}

As is seen from (5) quantum effects do not shift the quantum melting line from the origin $P=0, T=0$, but its slope should differ from the classical one at the same hard sphere diameter at any finite temperatures. As it follows from (3) and (5) for the slopes of the melting lines the following inequality should be valid,
$$(dP/dT)_{classical} > (dP/dT)_{quantum}\eqno(6)$$  
or
$$(dT/dP)_{classical} < (dT/dP)_{quantum}\eqno(7)$$.
On the other hand, the quantum correction to the hard sphere meting line should vanish at  temperature increasing as $T^{-1/2}$ (see (5)). The latter implies that the quantum melting line should be a curve, not a straight line as it is in case of classical melting. Moreover, the classical limit in the hard sphere quantum system can be reached only at $T\rightarrow \infty$.
Now we turn to the Quantum Monte – Carlo data describing melting of Boltzmann hard sphere system~\cite{runge, sese}. The Quantum Monte – Carlo calculations~\cite{runge, sese} were performed in the framework of Boltzmann statistics and for the particle’s diameter $\sigma=2.2$ \AA ~  and mass $m=4$. The corresponding data together with the classical hard sphere melting line are displayed in Fig.~\ref{fig2}. Consistently, calculations of the classical melting line were made for a case $\sigma=2.2$\AA. Note that the slopes of the melting lines in Fig.~\ref{fig2} agree with the prediction (6,7).

The Monte-Carlo data for the melting line of quantum hard sphere system and the classical melting line (see Fig.\ref{fig2}) can be well described with the expressions (8) and 
(9).
$$T_{q}=1.812\cdot 10^{-2}P^ {0.92202}\eqno(8)$$
$$T_{cl}=6.57\cdot 10^{-3} P\eqno(9)$$
As follows from (8) and (9) the both melting lines in Fig.\ref{fig2} can cross only at the coordinate origin $T=0, P=0$, implying that the classical limit is not reachable at any finite temperatures. 
In conclusion is worth noticing that inverting expression (8) one may obtain a melting line in a form similar to that of the soft sphere melting \cite{hoover, stishov},
$$P\sim T^{1+3/n}\eqno(10),$$ 
where $n\approx35$. Hence, one can describe effective quantum repulsion, appeared to be very steep, by  the relation $\Phi(r)\sim 1/r^{35}$ (compare it with the van Der Waals repulsion, where $n\approx12)$).

Finally, melting of a Boltzmann quantum system of hard spheres has been considered. It has been found that the quantum melting line always different from the classical line with exception for $T=0, P=0$, where the both lines crossed. The quantum corrections to the hard sphere meting line vanish with temperature as $T^{-1/2}$. So, the quantum melting line is a curve, not a straight line as it is in case of classical melting. The classical limit in the hard sphere quantum system cannot be reached at any finite temperature~\cite{stishov2}. At last one should note that the present analysis may serve as a validation of the simple approach ~\cite{jan} to the quantum hard sphere physics.


\begin{thebibliography}{99}
\bibitem{stishov2} So, the earlier analysis performed in Ref.\cite{stishov3, stishov4} appeared to be not quite correct in respect of quantum hard spheres melting.
\end{thebibliography}
\end{document}